\documentclass[useAMS,usenatbib]{mn2e}
\usepackage{graphicx}
\usepackage{epstopdf}
\usepackage{amsmath}
\usepackage{amssymb}
\title[Gas flows to an AGN]{Gas flows in an Active galactic nuclei. I. Two-phase gas inflow}
\author[Bu, Yang \& Zhu]{De-Fu Bu$^1$\thanks{E-mail: dfbu@shao.ac.cn}, Xiao-Hong Yang$^2$\thanks{E-mail: yangxh@cqu.edu.cn} and Bocheng Zhu$^1$ \\
$^{1}$Key Laboratory for Research in Galaxies and Cosmology, Shanghai Astronomical Observatory, Chinese Academy of Sciences,\\ 80 Nandan Road, Shanghai 200030, China \\
$^2$ Department of Physics, Chongqing University, Chongqing 400044, China}
\begin{document}

\pagerange{\pageref{firstpage}--\pageref{lastpage}} \pubyear{2002}

\maketitle

\label{firstpage}

\begin{abstract}
We perform two-dimensional hydrodynamic simulations to study the non-rotating gas flow in the region from 1-1000 parsec. This region connects the galaxy scale and the accretion disk scale. The gas is irradiated by the photons generated by the accretion disk system near the central black hole with $10^8$ solar mass. We assume that the luminosity of the central accretion system is constant and equals to $0.3$ Eddington luminosity. Gas with density higher than $10^{-24} {\rm g \ cm^{-3}}$ is found to be thermally unstable. Two phase, cold and hot, gas flow will form due to thermal instability. We calculated the ratio of cold gas accretion rate to hot gas accretion rate. This ratio is a function of gas density and generally increases with the increase of gas density. The maximum value of this ratio is $16$, when gas density is $10^{-21} {\rm g \ cm^{-3}} $. Gas with density lower than $10^{-24} {\rm g \ cm^{-3}}$ is found to be thermally stable and the gas flow is one-phase. The applications of the results are briefly discussed.
\end{abstract}


\section{Introduction}
It is generally believed that an active galactic nucleus (AGN) is composed by a super massive black hole at the center (see Heckman \& Best 2014 for reviews), a surrounding accretion disk and at larger scales broad line region (BLR) and narrow line region (NLR). The dynamics of accretion disk/flow have been studied extensively by numerical simulations (e.g., Stone et al. 1999; McKinney \& Gammie 2002; Yuan et al. 2012, 2015; Narayan et al. 2012).

In this paper, we focus on the region from $\sim 1 \ {\rm parsec}$ to $1000 \ {\rm parsec}$, which is roughly the NLR. There are several motivations for the present paper. The first motivation is about the gas supply to the accretion disk/flow located at smaller scale. As introduced above, there are many works studying the accretion disk/flow. In these simulations, the outer boundary is usually located at several hundreds/thousands Schwarzschild radius. The gas reservoir to form an accretion disk/flow is put by hand at hundreds/thousands Schwarzschild radius.  Actually, the gas to form the central accretion disk/flow comes from parsec scale. How does the gas fall from the parsec scale to the accretion disk/flow scale is poorly studied. In order to study the gas infall from parsec scale to hundreds/thousands Schwarzschild radius, we need ``proper" boundary conditions for gas at parsec scale. The inner boundary of the simulations in this paper is located at 1 parsec. Therefore, the results in this paper can be used as the boundary conditions to study gas infall from parsec to accretion disk/flow scale.

\begin{figure*}
\begin{center}
\includegraphics[scale=0.5]{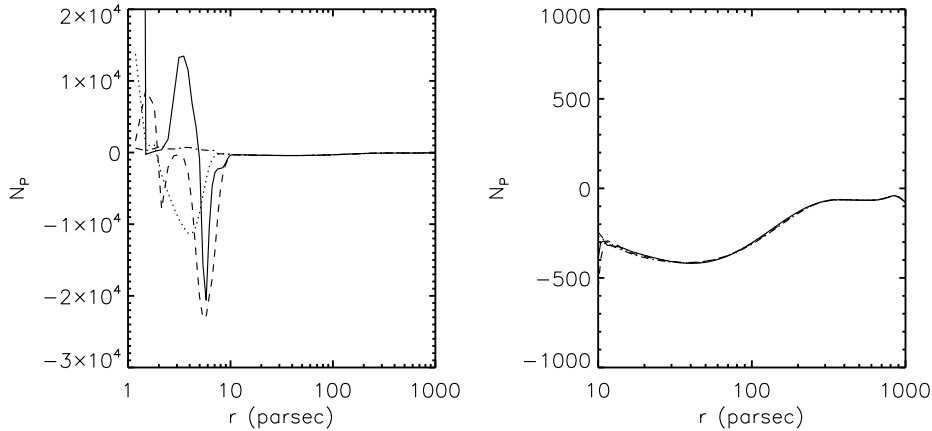}\hspace*{0.7cm} \\
\hspace*{0.5cm} \caption{Radial distributions of $N_P=\partial \zeta /\partial T |_p$ at $5.15\times 10^7$ year for model D21T5. The solid, dotted, dashed and dotted-dashed lines corresponds to $\theta = 0^\circ$, $30^\circ$, $60^\circ$ and $90^\circ$, respectively. The right panel is a vertical axis zoom in for the region $10-1000$ parsec. At this moment, we can see that in most of the radial computational domain, $N_P$ is negative. We note that for other $\theta$ angles and for other snapshots, we also find that $N_P < 0$ when $r>6$ parsec.
\label{Fig:Np}}
\end{center}
\end{figure*}

The second motivation is to study how and under what conditions can cold clumps (or filaments) form in the NLR. The kinematic studies of the cloud in the NLRs in some nearby Seyfert galaxies have shown that cold clumpy outflows are present (Crenshaw \& Kraemer 2000; Ruiz et al. 2001; Das et al. 2005, 2006; Crenshaw et al. 2010; Fischer et al. 2010, 2011). The outflow is accelerated from $v=0$ at $r \sim 0 \ {\rm parsec}$ to $\sim 1000 \ {\rm km \cdot s^{-1}}$ at one hundred to several hundreds parsecs. At even larger radii, the velocity of outflow decreases linearly with increase of radius to 0. The mass outflow rate and kinetic power of the cold clumpy outflow also first increase and then decrease with radius. There are several works studying the acceleration mechanisms of the cold clumpy outflow in the NLRs (Das et al. 2007; Mou et al. 2017). In these works, it is assumed that cold clumpy gas is present. However, how the cold clumpy gas is formed has not been studied. In the present paper, we study how does the cold clumpy gas form in the NLRs. The radiation from the region very close to the central black hole can propagate through the NLR. The gas in the NLR can be ionized. The gas can also be heated up by Compton heating and photoionization heating. The cooling mechanisms of gas include recombination cooling, line cooling and bremsstrahlung cooling. Under some conditions, the gas in this region is thermally unstable. The consequence of thermal instability is that two phase (hot and cold) medium can form and exist (Mo\'scibrodzka \& Proga 2013; Proga \& Waters 2015; Waters \& Proga 2019). We study under what gas conditions, can two phase medium form in the NLR.

Mo\'scibrodzka \& Proga (2013, hereafter MP13) studied the thermal instability of gas in the NLRs. There are big differences between MP13 and the present work. First, in MP13, it is assumed that there is only X-ray radiated out from the central region very close to the black hole. MP13 only considers the X-ray heating/cooling processes and the radiation pressure due to X-rays is neglected. In reality, most of the photons emitted by an AGN are in UV bands. In this paper, we consider more realistic case. We assume that $95\%$ of the photons produced by the central AGN are UV photons. The left $5\%$ of photons are X-rays. In addition to the radiative heating/cooling processes, we also consider the radiation pressure of the photons from the central region. Second, in MP13, the authors only focus on gas with initial density and temperature to be $10^{-23} \ {\rm g \ cm^{-3}}$ and $10^5$ K, respectively. Therefore, the results in MP13 cannot tell when initial density and temperature change, what the results will change correspondingly. In the present paper, we perform simulations with different initial gas densities and temperatures. We study under what gas conditions, can two phase medium form.

Now, we introduce the third motivation of this paper. In some large scale numerical simulations studying galaxy formation (Di Matteo et al. 2008; Dubois et al. 2010), the Bondi radius cannot be resolved. In this case, it is impossible to directly calculate the accretion rate of the central black hole. In these works, in their sub-grid models of AGN feedback, the authors usually boot up the accretion rate by assuming that the central black hole accretion is dominated by an unresolved cold phase. The Bondi radius can be expressed as $R_{\rm B} \approx 150 (M_{\rm BH}/10^8M_{\odot})(T_\infty/10^5{\rm K})^{-1}$ parsec, with $M_{\rm BH}$, $M_\odot$ and $T_\infty$ being black hole mass, solar mass and gas temperature at infinity, respectively. According to the parameters adopted in this paper, we can well resolve the Bondi radius. In this paper, we will show that under what conditions, can cold phase accretion be important. We will also give the ratio of cold gas accretion rate to hot gas accretion rate as a function of gas properties. These results in this paper can be used in the sub-grid models of AGN feedback in these kinds of galaxy formation simulations.

The structure of this paper is as follows. In Section 2, we present the equations, numerical settings, physical assumptions and initial and boundary conditions of the simulations. In Section 3, we present our results. We provide the discussions in Section 4. In Section 5, we introduce the applications of our results. We summarize our results in Section 6.

\section{Numerical method }
We set the mass of the central black hole to be $M_{\rm BH}=10^8M_{\odot}$. We adopt a simple physical set up. We assume that the angular momentum of the gas is zero. To keep the problem as simple as possible, we assume that the luminosity of the black hole accretion disk/flow, which is not resolved, is fixed. We assume that the bolometric luminosity of the black hole accretion disk/flow is $L_{\rm bol}=0.3L_{\rm Edd}$, with $L_{\rm Edd}$ being Eddington luminosity. We do not calculate the luminosity of the black hole accretion disk/flow based on the accretion rate at the inner radial boundary. The reasons are as follows. Our inner radial boundary is located at 1 parsec, which is far from the inner most stable orbit where most of the radiation is produced. In order to calculate the luminosity, we first need to assume the accretion mode of the accretion disk/flow. Second, in order to calculate the luminosity self-consistently, we also need to assume a radiative efficiency. Both of the two steps will complicate the problem.

\begin{figure}
\begin{center}
\includegraphics[scale=0.45]{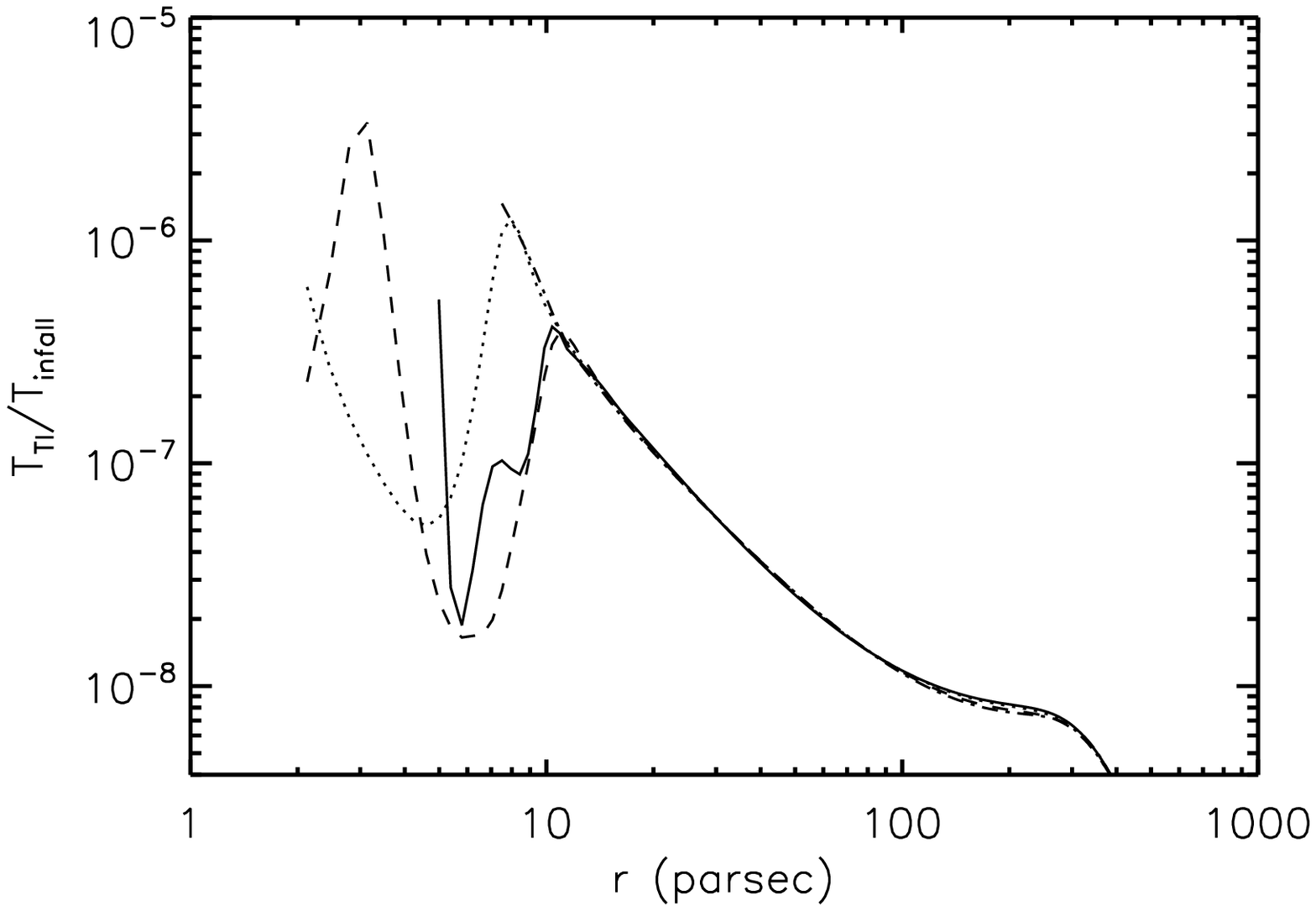}\hspace*{0.7cm}
\hspace*{0.5cm} \caption{Radial distributions of the ratio of growth timescale of thermal instability to the gas infall timescale in the region $N_P < 0$ at $5.15\times 10^7$ year for model D21T5. The solid, dotted, dashed and dotted-dashed lines corresponds to $\theta = 0^\circ$, $30^\circ$, $60^\circ$ and $90^\circ$, respectively. We note that for other $\theta$ angles and for other snapshots, we also find that $T_{\rm TI}/T_{\rm infall} \ll 1$. \label{Fig:timescaleD21T5}}
\end{center}
\end{figure}

We use the ZEUS-MP code (Hayes et al. 2006) to perform two-dimensional simulations in spherical coordinates ($r$,$\theta$,$\phi$). The hydrodynamic equations solved are as follows,
\begin{equation}
 \frac{d\rho}{dt} + \rho \nabla \cdot {\bf v} = 0,
\end{equation}
\begin{equation}
 \rho \frac{d{\bf v}}{dt} = -\nabla p - \rho \nabla \Phi + \rho \bf{F_{rad}}
\end{equation}
\begin{equation}
 \rho \frac{d(e/\rho)}{dt} = -p\nabla \cdot {\bf v} + \rho \dot E
\end{equation}
In these equations, $\rho$ is density, $\bf v$ is velocity, $e$ is internal energy per unit volume, $p$ is gas pressure, $\Phi$ is the gravitational potential of the central black hole, $\bf{F_{rad}}$ is the radiation pressure force due to Thomson scattering, $\dot E$ is the net heating/cooling rate per unit mass. In order to close the equations, we adopt ideal gas equation of state $p=(\gamma-1)e$, with $\gamma=5/3$.

For the accretion disk/flow model in an AGN, it is generally believed that a cold thin disk at the midplane is surrounding the black hole. Above and below the thin disk, there is a compact hot corona (Reis \& Miller 2013; Uttley et al. 2014; Chainakun et al. 2019). The cold thin disk mainly emits photons in UV band. The compact hot corona emits X-ray photons. We assume that the cold thin disk emits $95\%$ of the bolometric luminosity in UV band. $5\%$ of the bolometric luminosity is in X-ray band and is emitted by the compact corona. Oservations to some luminous AGNs show that the X-rays only contributes $5\%$ or less to the total luminosity. In Liu et al. (2013), it is also assume that X-ray contributes $5\%$ of the total luminosity.  As in previous works, we assume that the X-ray photons are emitted spherically (Proga et al. 2000; Liu et al. 2013), with the X-ray flux to be $F_{\rm X}=0.05 L_{\rm bol} \exp{(-\tau_{\rm X})}/4\pi r^2$. $\tau_{\rm X}=\int_0^r \rho \kappa_{\rm es} dr$ is the X-ray scattering optical depth, with $\kappa_{\rm es}= 0.4 \ {\rm cm^2 \ g^{-1}}$ being the Thomson scattering opacity. As done in Proga et al. (2000), we assume that the flux of UV photons is $F_{\rm UV}=2f_{\rm UV}\cos(\theta)L_{\rm bol}\exp(-\tau_{\rm UV})/4\pi r^2$, with $f_{\rm UV}=0.95$ being the ratio of UV luminosity to the bolometric luminosity. The flux of UV photons is maximum at the pole ($\theta=0$), it decreases towards the midplane ($\theta=\pi/2$). At the midplane, the flux of UV photons is zero. In this paper, we set $\tau_{\rm UV}=\tau_{\rm X}$.

\begin{figure*}
\begin{center}
\includegraphics[scale=0.5]{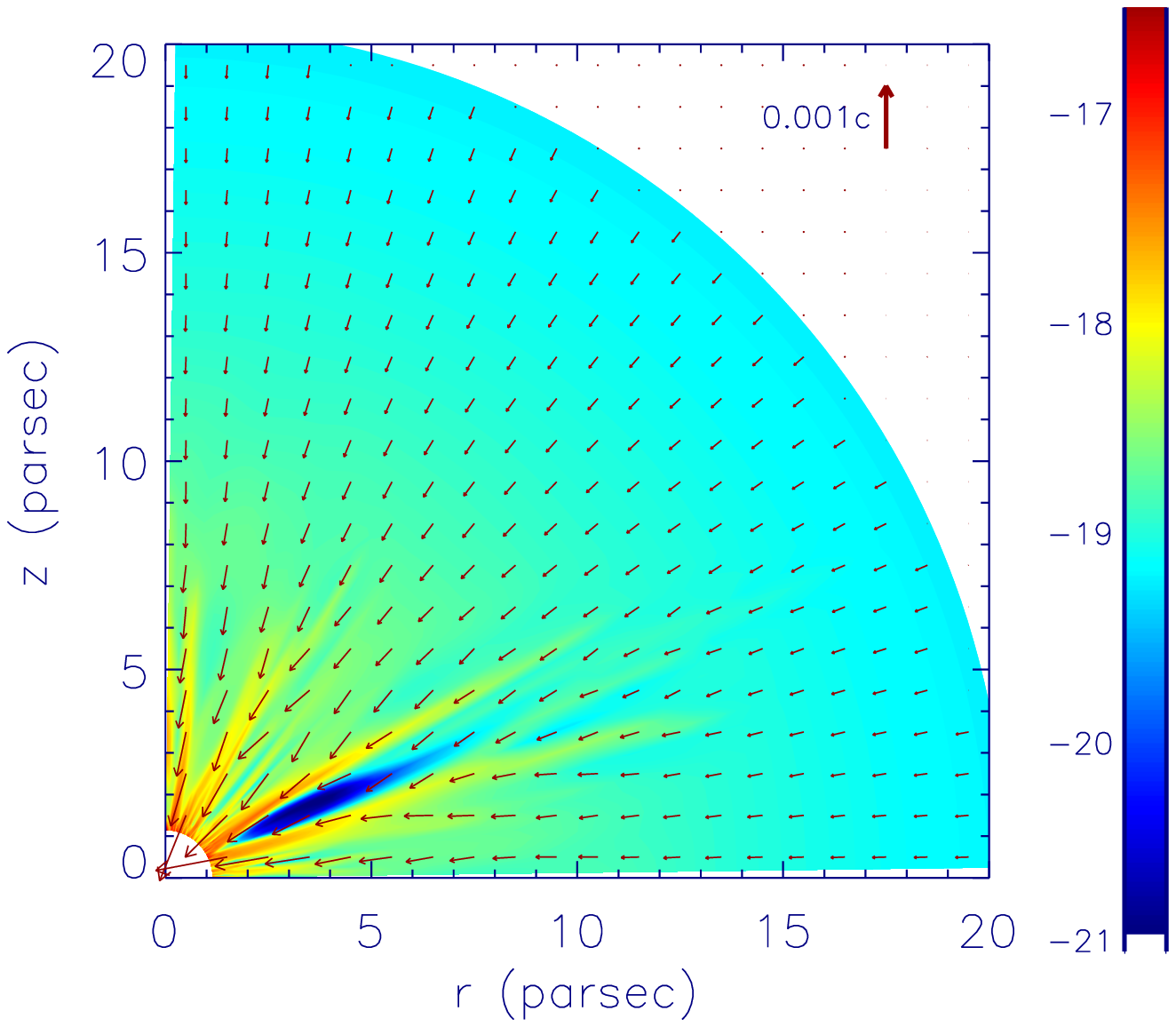}\hspace*{0.1cm}
\includegraphics[scale=0.5]{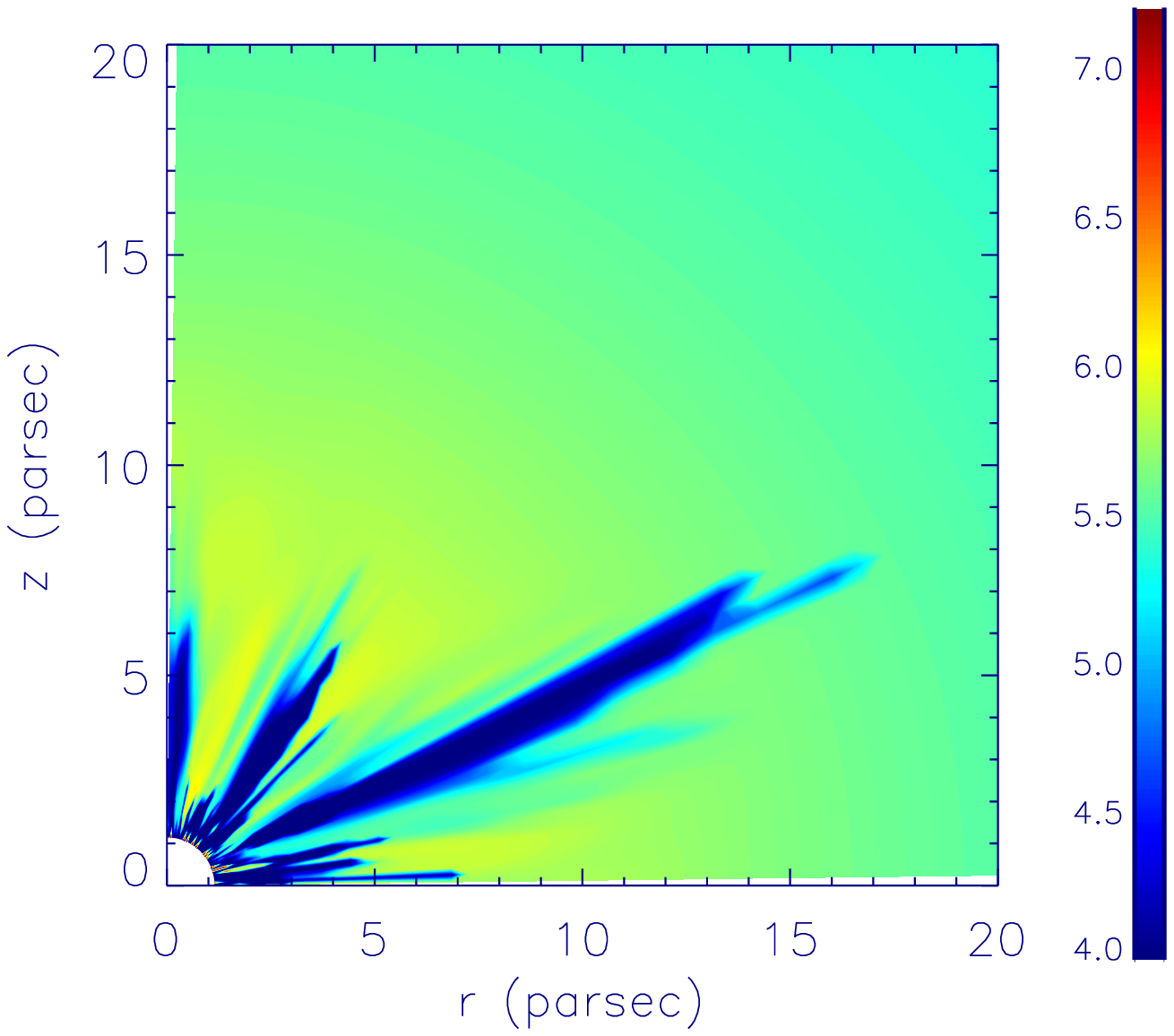}\hspace*{0.7cm}\\
\hspace*{0.5cm} \caption{Two-dimensional properties at $5.15\times 10^7$ year for model D21T5. The left panel plots logarithm density (color) and velocity vector (arrows). The right panel plots logarithm temperature (color). \label{Fig:vectorD21T5}}
\end{center}
\end{figure*}

Before introducing cooling/heating functions, we first define the ionization parameter as follows,
\begin{equation}
\xi=\frac{4\pi F_X}{n}=0.05 L_{\rm bol} \exp(-\tau_X)/n r^2
\end{equation}
$n=\rho/\mu m_p$ is gas number density, with $\mu=1$ and $m_p$ being mean molecular weight and proton mass, respectively.

\begin{table*} \caption{Models and results }
\setlength{\tabcolsep}{4mm}{
\begin{tabular}{cccccccc}
\hline \hline
 Model &  $\rho_0$ & $T_0$ & $\tau$ & $R_{\tau=1}$ & Cold clumps   & $\dot M_{\rm C}/\dot M_{\rm H}$ & $\dot M_{\rm C}$ \\

  &   ($10^{-21}\text{g cm}^{-3}$) & $ 10^5 {\rm K}$ &  &  &      &  &($M_\odot \ {\rm year^{-1}}$)   \\
(1)             & (2)          & (3)         &  (4)  & (5)     &     (6)          &        (7)   & (8)   \\

\hline\noalign{\smallskip}
D21T5      & 1 & 1 & 14.9 & 1.2 parsec &  Yes & $16 $ & $1016$  \\
D22T5      & 0.1 & 1 & 11.7 & 1.2 parsec & Yes & $14 $ & $906$  \\
D23T5      & 0.01 & 1 & 0.46 & - & Yes & $ 2.2 $ & $33$  \\
D24T5      & 0.001 & 1 & 0.0017 & - & No  & $ 0 $ & $0$  \\
D21T6      & 1 & 10 & 13.4 & 1.26 parsec & Yes  & $ 14 $ & $903$  \\
D22T6      & 0.1 & 10 & 7.51 & 1.3 parsec & Yes  & $ 12 $ & $567$  \\
D23T6      & 0.01 & 10 & 2.44 & 1.5 parsec & Yes  & $ 5 $ & $600$  \\
D24T6      & 0.001 & 10 & 1.37 & 3.0 parsec & Yes  & $ 5 $ & $503$  \\
D25T6      & 0.0001 & 10 & 0.00015 & - & No  & $ 0 $ & $0$  \\
D21T7      & 1 & 100 & 13.4 & 1.2 parsec & Yes  & $ 14 $ & $831$  \\
D22T7      & 0.1 & 100 & 2.1 & 3.6 parsec & Yes  & $ 7 $ & $149$  \\
D23T7      & 0.01 & 100 & 10.2 & 1.2 parsec &Yes  & $ 13 $ & $974$  \\
D24T7      & 0.001 & 100 & 0.88 & - & Yes  & $ 5 $ & $163$  \\
D25T7      & 0.0001 & 100 & 0.00017 & - & No  & $ 0 $ & $0$  \\
\hline\noalign{\smallskip}
\end{tabular}}

Note: Col. 1: model names. Col 2: the density for initial condition. Col 3. the temperature for initial condition. Col 4. the time-averaged radial optical depth along $\theta=45^\circ$, the optical depth is obtained by integration from the inner to the outer radial boundaries. Col 5. the radius at which the optical depth along $\theta=45^\circ$ equals to 1.  Col. 6: Whether can cold streams/clumps ($T<10^5$ K) form. Col. 7: time-averaged ratio of cold gas accretion rate to hot gas accretion rate. Col. 8: time-averaged accretion rate for cold gas (with $T<10^5$ K).
\end{table*}

\begin{table*} \caption{Models and results }
\setlength{\tabcolsep}{4mm}{
\begin{tabular}{cccccccc}
\hline \hline
 Model &  $\rho_0$ & $T_0$ & $\tau$ & $R_{\tau=1}$ & Cold clumps   & $\dot M_{\rm C}/\dot M_{\rm H}$ & $\dot M_{\rm C}$ \\

  &   ($10^{-21}\text{g cm}^{-3}$) & $ 10^5 {\rm K}$ &  &  &      &  &($M_\odot \ {\rm year^{-1}}$)   \\
(1)             & (2)          & (3)         &  (4)  & (5)     &     (6)          &        (7)   & (8)   \\

\hline\noalign{\smallskip}
D21T5-XL      & 1 & 1 & 9.6 & 1.2 parsec &  Yes & $40 $ & $698$  \\
D21T5-XH      & 1 & 1 & 20 & 1.2 parsec &  Yes & $10 $ & $1215$  \\

\hline\noalign{\smallskip}
\end{tabular}}

Same as Table 1, but for the two test simulations D21T5-XL and D21T5-XH. The two test simulations are performed to compare with model D21T5. The UV flux in the two test simulations is set same as that in model D21T5.  In model D21T5-XL, the X-ray flux is 5 times lower than that in model D21T5. In model D21T5-XH, the X-ray flux is 5 times higher than that in model D21T5. The results are analyzed in Section 3.3.
\end{table*}

When gas is illuminated by the X-ray photons, it can be heated/cooled. The cooling/heating rates of the X-ray irradiated gas used in this paper are same as those in Proga et al. (2000). The cooling/heating rates are for gas with cosmic abundances. Blondin (1994) studied the cooling/heating rates for gas illuminated by a 10 kev bremsstrahlung spectrum. It is found that it is possible to fit analytical formulae to the heating/cooling rates obtained from detailed photonization calculations. The expressions for heating/cooling rates in this paper are by using Blondin¡¯s (1994) results. The gas can be heated/cooled by Compton process. The gas can also be heated/cooled by the photoionization heating-recombination cooling process. The Compton heating/cooling rate is:
\begin{equation}
S\rm c=8.9\times10^{-36}n^2(T_X-4T)\xi \ {\rm erg \ cm^{-3} \ s^{-1}}
\end{equation}
In this equation, $T_X$ is the Compton temperature of the X-ray photons, $T$ is the temperature of the accreting gas. As done by MP13, we assume that $T_X=1.16\times 10^8$ K.
The sum of photoionization heating-recombination cooling rate is:
\begin{equation}
G\rm x=1.5\times10^{-21}n^2 \xi^{1/4} T^{-1/2} (1-T/T_X) \ {\rm erg \ cm^{-3} \ s^{-1}}
\end{equation}
The gas can also cool by the bremsstrahlung cooling,
\begin{equation}
B\rm r=3.3\times 10^{-27}n^2\sqrt{T} \ {\rm erg \ cm^{-3} \ s^{-1}}
\end{equation}
and the line cooling,
\begin{equation}
\begin{split}
L_{\rm line}=1.7\times10^{-18}n^2\exp(-1.3\times10^5/T)/\xi/\sqrt{T}+\\
            10^{-24} n^2 \ {\rm erg \ cm^{-3} \ s^{-1}}
\end{split}
\end{equation}
In Equation (3), the net heating/cooling rate $\rho \dot E=S {\rm c} + G {\rm x} + B {\rm r} + L_{\rm line}$.

In this paper, the radiation pressure force due to electron scattering ($\bf{F_{rad}}$) is also included. Because the inner radial boundary of our simulations is significantly larger than the region where most of the photons of an AGN are emitted out, we can treat the AGN emitting region as a point source.
The radiation force only has the radial component,
\begin{equation}
{F_{\rm rad, r}}=\frac{\kappa_{es}}{c}\left[F_{\rm X}+F_{\rm UV}\right]
\end{equation}

\subsection{Initial and boundary conditions}
Initially, gas with uniform density ($\rho_0$) and temperature ($T_0$) is put in the whole computational domain. Our radial computation domain is $1 \ {\rm parsec} \leq r \leq 1000 \ {\rm parsecs}$. In $\theta$ direction, we have the domain $0 \leq \theta \leq \pi/2$. Our resolution is $192 \times 64$. In $r$ direction, we adopt logarithm grids in order to well resolve the inner region. In $\theta$ direction, grids are uniformly spaced. At the inner radial boundary, we use outflow boundary conditions. At the outer radial boundary, we set the boundary conditions as follows. When gas has $v_r < 0 $ at the last active zone at a given $\theta$ angle, then at this angle, we inject gas into the computational domain. The density, temperature, of the injected gas are equal to those for the initial condition. If we have $v_r>0$ at the last active zone at a given $\theta$ angle, then at this angle, we employ outflow boundary conditions. This kind of outer boundary conditions can continuously supply gas at the outer boundary. At $\theta=0$, we use axis-of-symmetry boundary conditions. At $\theta=\pi/2$, reflecting boundary conditions are employed.

\section{Results}
\begin{figure}
\begin{center}
\includegraphics[scale=0.5]{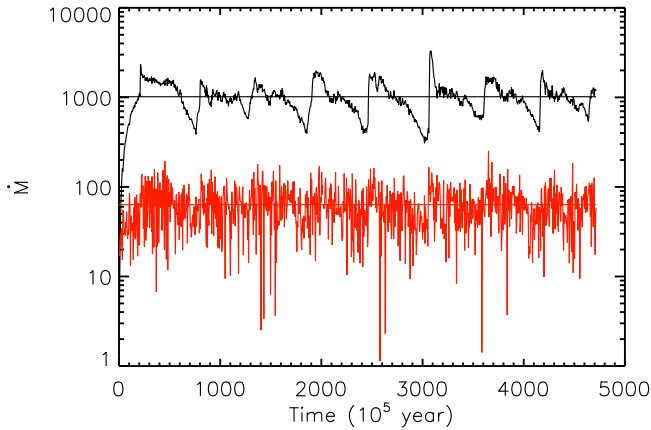}\hspace*{0.7cm}
\hspace*{0.5cm} \caption{Time evolution of the mass accretion rate at the inner boundary for model D21T5. The accretion rate is in unit of $M_\odot \ {\rm year^{-1}}$. The fluctuating black and red lines correspond to the cold gas accretion rate and hot gas accretion rate, respectively. The horizontal black and red lines correspond to the time-averaged values of cold gas accretion rate and hot gas accretion rate, respectively. \label{Fig:mdotD21T5}}
\end{center}
\end{figure}

We summarize all the models in Table 1. In this table, the second and third columns give the density and temperature values for initial condition and the outer boundary condition if gas flows inwards at the outer boundary. The fourth column gives the time-averaged radial optical depth along $\theta=45^\circ$, the optical depth is obtained by integration from the inner to the outer radial boundaries. The fifth column gives the radius at which the optical depth along $\theta=45^\circ$ equals to 1. The sixth column gives the information whether cold gas streams/clumps (with $T<10^5$ K) can form. The seventh column gives the time-averaged ratio of cold gas accretion rate to hot gas accretion rate. The eighth column gives the time-averaged accretion rate for cold gas in unit of $M_\odot \cdot \ {\rm year^{-1}}$.

\subsection{Models with $T_0=10^5$ K}
The linear stability analysis done by Field (1965) has shown that the gas illuminated by X-ray is thermally unstable when
\begin{equation}
N_P=\partial \zeta /\partial T |_p < 0
\end{equation} where $\zeta$ is the cooling function. In this paper, we have $\zeta = -\dot E$ (see Equation 3 for the definition of $\dot E$). The thermal instability of gas illuminated by X-rays is also studied by MP13 and Waters \& Proga (2019).

We first analyze model D21T5 with $\rho_0=10^{-21}{\rm g/cm^3}$ and $T_0=10^5{\rm K}$. In Figure \ref{Fig:Np}, we show the snapshot of the radial distribution of $N_P$ at $t = 5.15 \times 10^7$ year when there is a peak in the mass accretion rate of cold phase gas ($T < 10^5$ K, see the information below). In this figure, the solid, dotted, dashed and dotted-dashed lines corresponds to $\theta = 0^\circ$, $30^\circ$, $60^\circ$ and $90^\circ$, respectively. As can be seen, in most of the computational domain ($r> 6$ parsec), $N_P < 0$, the gas is thermally unstable. We point out that at other $\theta$ angles, the results are quite similar that in the region ($r > 6$ parsec), $N_P<0$. We have also find that for the accretion flow at any other time, in most of the computational domain ($r > 6$ parsec), $N_P < 0$. When the growth timescale of thermal instability is shorter than the gas infall timescale ($T_{\rm infall} = r/v_r$), thermal instability can grow. The growth timescale of thermal instability is (Field 1965),
\begin{equation}
T_{\rm TI} = \frac{1}{N_P/(2.5k/m_p)}
\end{equation}
where $k$ is the Boltzmann constant.
In Figure \ref{Fig:timescaleD21T5}, we show the snapshot of the ratio of thermal instability growth timescale to the gas infall timescale at $5.15\times 10^7$ year for model D21T5. The solid, dotted, dashed and dotted-dashed lines corresponds to $\theta = 0^\circ$, $30^\circ$, $60^\circ$ and $90^\circ$, respectively. From this figure, we can see that the thermal instability growth timescale is several orders of magnitude shorter than the gas infall timescale. We note that at other $\theta$ angles and for other snapshots, the results are similar. The thermal instability can grow.

The consequence of thermal instability is that the accretion gas will form two phase, hot and cold, medium. Following MP13, we define gas with $T < 10^5$ K as cold gas and gas with $T > 10^5$ K as hot gas. In Figure \ref{Fig:vectorD21T5}, we plot the two-dimensional properties at $5.15\times 10^7$ year for model D21T5. The left panel plots logarithm density (color) and velocity vector (arrows). The right panel plots logarithm temperature (color). It is clear that two-phase accretion flow forms. Both the low temperature (high density) and high temperature (low density) streams fall on to the center.

\begin{figure*}
\begin{center}
\includegraphics[scale=0.5]{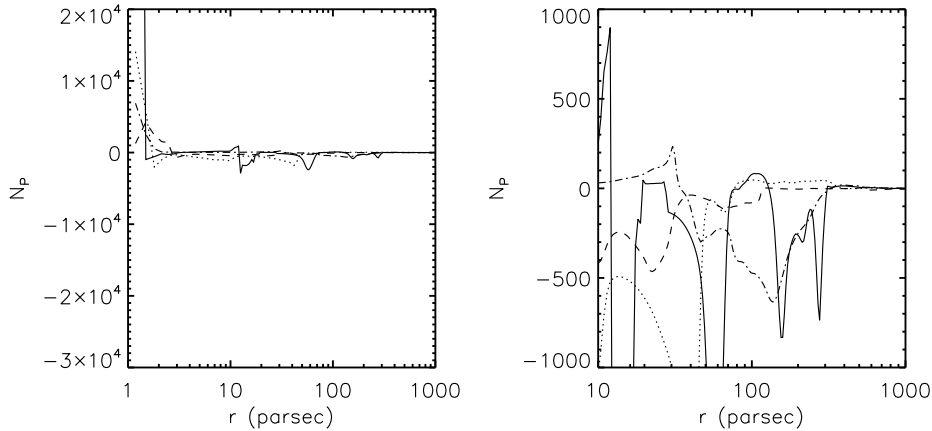}\hspace*{0.7cm} \\
\hspace*{0.5cm} \caption{Same as Figure \ref{Fig:Np}, but for model D22T5 at $t=7.65\times 10^7$ year.
\label{Fig:NpD22T5}}
\end{center}
\end{figure*}

\begin{figure}
\begin{center}
\includegraphics[scale=0.45]{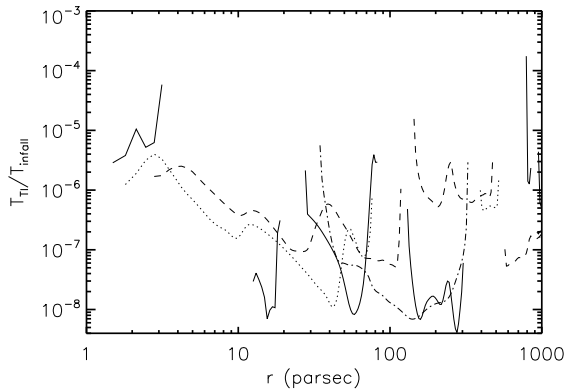}\hspace*{0.7cm}
\hspace*{0.5cm} \caption{Radial distributions of the ratio of growth timescale of thermal instability to the gas infall timescale in the region $N_P < 0$ at $7.65\times 10^7$ year for model D22T5. The solid, dotted, dashed and dotted-dashed line corresponds to $\theta = 0^\circ$, $30^\circ$, $60^\circ$ and $90^\circ$, respectively. We note that at other $\theta$ angles and for other snapshots, the results are similar. \label{Fig:timescaleD22T5}}
\end{center}
\end{figure}

Figure \ref{Fig:mdotD21T5} shows the time evolution of the mass accretion rate at the inner boundary. The accretion rate is in unit of $M_\odot \ {\rm year^{-1}}$. The fluctuating black and red lines correspond to the cold gas accretion rate and hot gas accretion rate, respectively. The horizontal black and red lines correspond to the time-averaged values of cold gas accretion rate and hot gas accretion rate, respectively.
As found by MP13, both cold gas accretion rate and hot gas accretion rate fluctuate with time. When the dense cold gas streams formed due to thermal instability across the inner boundary, we will observe a peak in the black solid line.
The time-averaged accretion rate of cold phase gas is 16 times of the time-averaged accretion rate of the hot phase gas.

Now, we study the dependence of the results on gas density. According to the cooling function $\zeta$, the cooling rate per unit mass is positively correlated with the gas density. Different cooling terms (Equations (5)-(8)) have different dependences on density. Gas with higher density is easier to cool to a much lower temperature. In other words, the thermal instability can be much easier to occur for gas with higher density.

In model D22T5, the initial gas density is 10 times smaller than that in model D21T5. Models D21T5 and D22T5 have same initial gas temperature. In Figure \ref{Fig:NpD22T5}, we plot the snapshot of the radial distribution of $N_P$ at $t = 7.65 \times 10^7$ year when there is a peak in the mass accretion rate of cold phase gas ($T < 10^5$ K, see the information below). In this figure, the solid, dotted, dashed and dotted-dashed lines correspond to $\theta = 0^\circ$, $30^\circ$, $60^\circ$ and $90^\circ$, respectively.
We note that at other $\theta$ angles and for other snapshots, the results are similar. Comparing Figures \ref{Fig:Np} and \ref{Fig:NpD22T5}, we can find the following differences. First, in model D21T5, at any radii in the region $r>10$ parsec, $N_P$ is negative, the gas is thermally unstable. However, in model D22T5, in the region $r>10$ parsec, $N_p$ can be either positive or negative. In other words, in model D21T5, in the region $r>10$ parsec, thermal instability can occur at any location. However, in model D22T5, thermal instability can only occur at some radii. Second, in the thermally unstable region ($N_P<0$) at radius $r<10$ parsec, the growth timescale of thermal instability in model D21T5 is shorter (see Equation 11 for the calculation of $T_{TI}$).

In Figure \ref{Fig:timescaleD22T5}, we show the radial distribution of the ratio of growth timescale of thermal instability to the gas infall timescale in the region $N_P < 0$ at $7.65\times 10^7$ year for model D22T5. In this figure, the solid, dotted, dashed and dotted-dashed lines correspond to $\theta = 0^\circ$, $30^\circ$, $60^\circ$ and $90^\circ$, respectively. We note that at other $\theta$ angles and for other snapshots, the results are similar. The lines are discontinuous. This is because we just show the results for the region with $N_P < 0$. As can been seen that in the thermally unstable region, the growth timescale of thermal instability is significantly shorter than the gas infall timescale. In this model, two phase gas can form.

\begin{figure}
\begin{center}
\includegraphics[scale=0.5]{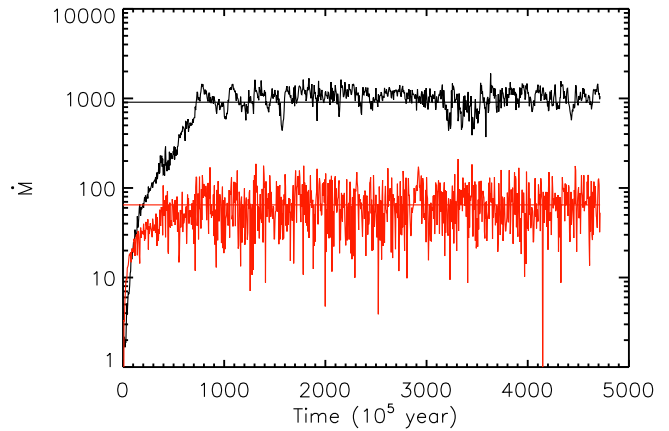}\hspace*{0.7cm}
\hspace*{0.5cm} \caption{Same as Figure \ref{Fig:mdotD21T5}, but for model D22T5. \label{Fig:mdotD22T5}}
\end{center}
\end{figure}

Figure \ref{Fig:mdotD22T5} shows the time evolution of mass accretion rate for both cold gas (black line) and hot gas (red line) measured at the inner boundary. Same as those in model D21T5, the mass accretion rates for both cold and hot gas fluctuate with time. Comparing Figures \ref{Fig:mdotD21T5} and \ref{Fig:mdotD22T5}, we can see that the amplitude of fluctuation for the cold gas accretion rate in model D22T5 is smaller. The amplitudes of fluctuation for hot gas accretion rates in the two models are similar. The time averaged cold gas accretion rate in model D22T5 is slightly smaller. Lower density in model D22T5 makes the cooling rate per unit mass in this model smaller. Therefore, compared to model D21T5, it is harder to form dense cold gas in model D22T5. Therefore, the cold gas accretion rate in model D22T5 is smaller. The time averaged hot gas accretion rate in model D22T5 is slightly higher than that in model D21T5. In model D22T5, the ratio of time averaged cold gas accretion rate to hot gas accretion rate is $14$. In model D21T5, this value is $16$.

\begin{figure}
\begin{center}
\includegraphics[scale=0.5]{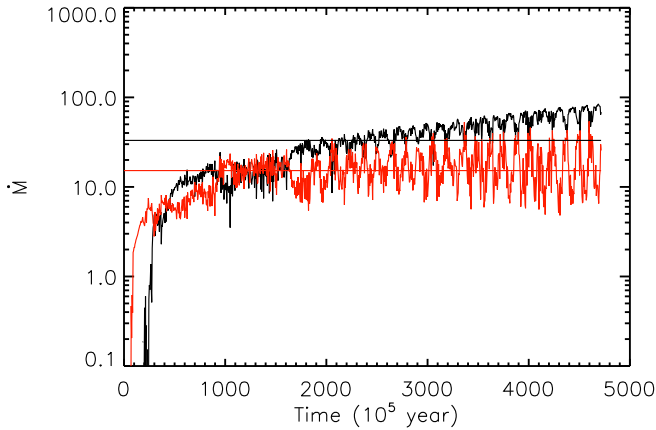}\hspace*{0.7cm}
\hspace*{0.5cm} \caption{Same as Figure \ref{Fig:mdotD21T5}, but for model D23T5. \label{Fig:mdotD23T5}}
\end{center}
\end{figure}

\begin{figure*}
\begin{center}
\includegraphics[scale=0.5]{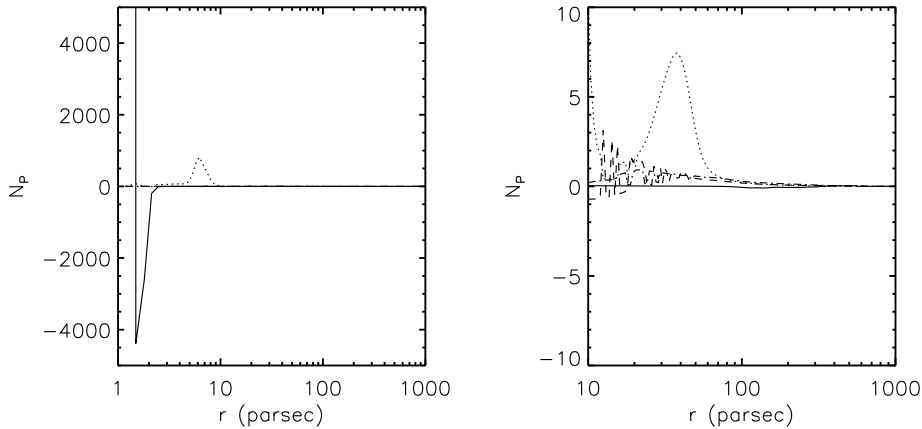}\hspace*{0.7cm} \\
\hspace*{0.5cm} \caption{Same as Figure \ref{Fig:Np}, but for model D23T5 at $t=4.6\times 10^8$ year.
\label{Fig:NpD23T5}}
\end{center}
\end{figure*}

\begin{figure}
\begin{center}
\includegraphics[scale=0.45]{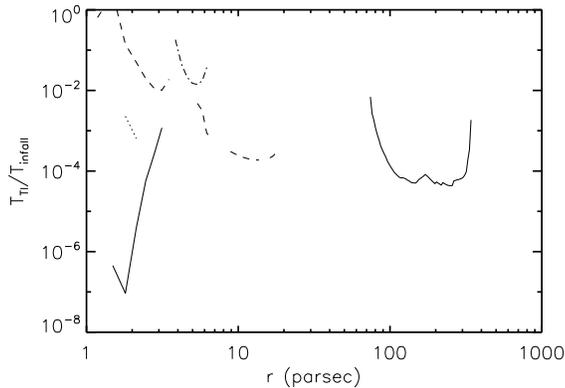}\hspace*{0.7cm}
\hspace*{0.5cm} \caption{Radial distributions of the ratio of growth timescale of thermal instability to the gas infall timescale in the region $N_P < 0$ at $4.6\times 10^8$ year for model D23T5. The solid, dotted, dashed and dotted-dashed lines corresponds to $\theta = 0^\circ$, $30^\circ$, $60^\circ$ and $90^\circ$, respectively. We note that at other $\theta$ angles and for other snapshots, the results are similar.  \label{Fig:timescaleD23T5}}
\end{center}
\end{figure}

In model D23T5, the value of $\rho_0 = 10^{-23} {\rm g \ cm^{-3}}$, which is 10 times smaller than that for model D22T5. For this model, we first show the time evolution of mass accretion rate in Figure \ref{Fig:mdotD23T5}. It can be seen that when time $< 1.6\times 10^8$ year, the cold gas accretion rate is roughly equal to the hot gas accretion rate. After $ 1.6\times 10^8$ year, the cold gas accretion rate gradually increases with time, and cold gas accretion rate becomes larger than the hot gas accretion rate. In this model, the gas density is much lower. The lower gas density has two consequences. First, the region where $N_P < 0$ is smaller compared to models D21T5 and D22T5 (see Figure \ref{Fig:NpD23T5}). Second, the growth timescale of thermal instability becomes longer (see Figure \ref{Fig:timescaleD23T5}). Compared to models D21T5 and D22T5, it takes longer time for the thermal instability to grow in model D23T5. This is the reason why in model D23T5, only after $1.6 \times 10^8 $year, can the cold gas accretion rate exceeds the hot gas accretion rate.

In model D24T5, the initial density is 10 times lower than that in model D23T5. We find that in model D24T5, the density is too low that the gas is thermally stable ($N_P > 0$). Therefore, in this model, no cold phase gas forms and the cold gas accretion rate is zero.

\subsection{Models with $T_0=10^6$ K and $T_0=10^7$ K}
For the models with $T_0=10^6$ K and $T_0=10^7$ K, our main findings about the two-phase accretion are very similar as those in models with $T_0=10^5K$. The findings are as follows (see also Table 1). When $\rho_0 \geq 10^{-24} \ {\rm g \ cm^{-3}}$, thermal instability can occur and cold phase gas can form. The ratio of time-averaged cold gas accretion rate to hot gas accretion rate is a function of $\rho_0$ (see Table 1 for detailed values). Generally, this ratio increases with the increase of $\rho_0$. When $\rho_0=10^{-21} \ {\rm g \ cm^{-3}}$, this ratio reaches its maximum value of 16.

\subsection{Effects of changing X-ray flux}
In order to study the effects of X-ray flux, we perform two test simulations D21T5-XL and D21T5-XH. The UV flux in the two test simulations is set same as that in model D21T5.  In model D21T5-XL, the X-ray flux is 5 times lower than that in model D21T5. In model D21T5-XH, the X-ray flux is 5 times higher than that in model D21T5. The results of the two test simulations are listed in Table 2. We calculated the ratio of cold gas accretion rate to hot gas accretion rate. In model D21T5-XL, this ratio is 40. In model D21T5-XH, this ratio is 10. In model D21T5, this ratio equals to 16. With the increase of X-ray flux, less gas will be accreted in cold mode. This is easy to be understood. More X-ray flux can have stronger heating, more gas will tend to be in hot mode. In MP13, the authors also studied the effects of changing X-ray flux. It is also found that with the increase of X-ray flux, more gas will be accreted in hot mode.

\section{Discussions}
In this paper, we assume that the luminosity of the central engine is fixed. However, as previous works show that when the luminosity is higher than $\sim$ 0.01 times the Eddington luminosity, the accretion is unstable and the luminosity should vary with time (Ostriker et al. 1976; Cowie et al. 1978; Ciotti \& Ostriker 2001). If the luminosity changes with time, we would expect that the accretion rate of cold gas should oscillate periodically around some mean value with time. The reason is as follows. When the luminosity is low, the Compton/photoionization heating is weak, the gas can cool down. Then cold gas will fall to the center, the accretion rate becomes high and consequently the central engine will have a high luminosity. When the luminosity becomes high, the heating becomes stronger, then the accretion gas may be heated to the temperature higher than Virial temperature. The accretion rate and luminosity decrease. The loop will always be present. In this case, as we mentioned, we would expect the accretion rate of cold gas oscillates periodically. However, it is hard to expect qualitatively that whether the mean value of the cold gas accretion rate in a realistic case is higher or lower than the value obtained in this paper.

We discuss the possible influence of resolution on the size of the cold clumps formed by thermal instability. Koyama \& Inutsuka (2004) studied the thermal instability of interstellar medium. They found that in calculations without thermal conduction, the size of the cold cloud is roughly the size of numerical grid. If thermal conduction is included, the size of cold cloud can be invariant with resolution of calculations. In the simulations with thermal conduction, the ``Field length'' should be well resolved. The Field length is the characteristic length scale of thermal conduction.  The Field length is equal to $\sqrt{\kappa T/\rho \dot E}$, with $\kappa$ being the conductivity. The conductivity $\kappa = 2500\sqrt{T}$ erg cm$^{-1}$ K$^{-1}$ s$^{-1}$. We take our fiducial model D21T6 as an example to calculate the Field length. In this model, the initial gas temperature is $10^6$K. We assume that the cooling is bremsstrahlung. We find that the Field length is roughly $10^{15}$ cm, which is three orders of magnitude smaller than the inner boundary of our simulations. Therefore, in order to well resolve the Field length, our finest gird should be more than 3 orders of magnitude smaller than the inner boundary. Currently, it is quite hard to do such high resolution simulations. There is the possibility that if the size of the cold cloud is too small, the conduction may evaporate the cloud given that the conduction timescale is shorter than the gas infall timescale. In this case, the gas may be just one-phase. In future it is quite necessary to do high resolution simulations with the Field length resolved to study the thermal instability.

\section{Applications}

Previous works studying the NLRs in some nearby Seyfert galaxies have shown that cold clumpy outflows are present (Crenshaw \& Kraemer 2000; Ruiz et al. 2001; Das et al. 2005, 2006; Crenshaw et al. 2010; Fischer et al. 2010, 2011). The formation of the cold clumpy gas in the NRLs has not been fully understood. In this work, our results show that in the NLRs, cold clumps can form by thermal instability when gas density is higher than $10^{-24} {\rm g \ cm^{-3}}$. How the cold clumpy gas is pushed out to form the observed clumpy outflows is interesting. In future, we plan to study the acceleration mechanism of clumpy outflows in the NLRs to explain the observations.

AGN feedback is believed to play an important role in the galaxy formation and evolution (Fabian 2012). Both radiation and wind from an AGN can interact with its host galaxy (e.g., Ciotti \& Ostriker 2007; Ostriker et al. 2010; Ciotti et al. 2017; Yuan et al. 2018). The properties of AGN winds in these simulations are set according to the results of simulations of accretion flows. Usually, the outer boundary of the accretion flow simulations are located at several hundreds to thousand Schwarzschild radius. How the gas is accreted from parsec scale to hundreds/thousands Schwarzschild radius is unknown. AGN winds may also be generated when gas falls from parsec scale to hundreds/thousands Schwarzschild radius. Therefore, it is necessary to perform simulations to study the gas accretion and wind generation in the region from parsec to hundreds/thousands Schwarzschild radius. The winds generated from this region should also be taken into account in future simulations with AGN feedback. In order to study how the gas falls from parsec scale to the accretion flow, one needs to know the gas properties at parsec scale. Results in this paper can be used as outer boundary conditions in simulations studying gas infall from parsec scale to hundreds/thousands Schwarzschild radius.

In some simulations studying galaxy formation (Di Matteo et al. 2008; Dubois et al. 2010), the resolution is too low and it is hard to directly calculate the accretion rate of an AGN. In these simulations, in order to boost up the accretion rate obtained, it is usually assumed that the central black hole accretion is dominated by an unresolved cold phase. In this paper, we have shown that at parsec to kilo-parsec scale, only when the density of gas is higher than $10^{-24} {\rm g \ cm^{-3}}$, can thermal instability occur and consequently cold gas forms. When the gas density at this scale is smaller than $10^{-24} {\rm g \ cm^{-3}}$, no cold gas can form. We have also get the ratio of cold gas accretion rate to hot gas accretion rate as a function of initial gas density and temperature. These results can be used in future in the sub-grid models of AGN feedback in galaxy formation simulations.

\section{Summary}
Two-dimensional hydrodynamic simulations are performed to study the properties of gas flow irradiated by AGN photons in the region from 1-1000 parsec. We set the problem as simple as possible. We assume that the luminosity of the central accretion system is constant and equal to $0.3L_{\rm Edd}$. We also assume that the angular momentum of gas is zero. We find that when gas density is higher than $10^{-24} {\rm g \ cm^{-3}}$, the gas is thermally unstable. The consequence of thermal instability is that two phase, cold and hot, gas will form. We also study the accretion rate of the two phase gas. We find that the ratio of cold gas accretion rate to hot gas accretion rate is a function of gas density. Generally, this ratio increases with the increase of gas density. The maximum value of this ratio is $16$, when gas density is $10^{-21} {\rm g \ cm^{-3}} $. When gas density is smaller than $10^{-24} {\rm g \ cm^{-3}}$, the gas is thermally stable and no cold clumpy gas can form.

\section*{Acknowledgments}
D.-F. Bu is supported in part the Natural Science Foundation of China (grants 11773053,
11573051, 11633006 and 11661161012), and the Key
Research Program of Frontier Sciences of CAS (No. QYZDJSSW-
SYS008). X.-H. Yang is supported by Natural Science Foundation of China (grant 11973018) and Natural Science Foundation Project of CQ CSTC (grant cstc2019jcyj-msxmX0581). This work made use of the High Performance Computing Resource in the Core
Facility for Advanced Research Computing at Shanghai Astronomical
Observatory.

\section*{Data availability}
The data underlying this article will be shared on reasonable request to the corresponding author.

\end{document}